\documentclass[
reprint,
amsmath,amssymb,
aps,
prl,
]{revtex4-2}
\usepackage{soul}
\usepackage{xcolor}
\usepackage{float}
\usepackage{array}
\usepackage{mathtools}
\usepackage{multirow}
\usepackage{graphicx}
\usepackage{dcolumn}
\usepackage{bm}
\usepackage{upgreek}
\usepackage{lipsum, babel, xcolor}
\usepackage{amsmath, amssymb}
\usepackage{comment}
\usepackage{natbib}
\usepackage{physics}
\usepackage{hyperref}
\hypersetup{colorlinks=true,citecolor=blue}

\begin{document}


\title{Unveiling charge dynamics on the generation of high extinction wide pulse generation on thin -film lithium tantalate}
\author{Ayed Al Sayem\textsuperscript{$\dagger$}, Shiekh Zia Uddin\textsuperscript{$\dagger$}, Ting-Chen Hu, Tam Huynh, Bongjun Choi, Alaric Tate, Mark Cappuzzo, Rose Kopf, Mark Earnshaw}

\affiliation{%
  Nokia Bell Labs, NJ, USA $\mathrm{^{1}}$
}%
\date{\today}

\begin{abstract}
We experimentally demonstrate high-extinction optical pulse generation in thin-film lithium tantalate (TFLT), achieving a wide range of pulse widths with sharp rise and fall edges, without measurable distortion or long temporal tails. We further provide a direct one-to-one comparison with thin-film lithium niobate (TFLN) modulators under identical measurement conditions. Our results reveal that the distorted pulse response observed in TFLN originates from charge activation and transport, whereas the distortion-free response in TFLT is enabled by the substantially larger activation energy of defect-related charge carriers. This larger activation barrier suppresses leakage-current-induced charging dynamics, thereby enabling stable and distortion-free pulse generation. The successful generation of high-extinction, distortion-free optical pulses in TFLT can play a significant role in integrated quantum photonic technologies, particularly for qubit preparation, measurement, and fast feedback circuits.
\end{abstract}

\maketitle

\section{Introduction}

High-extinction electro-optic modulation and optical pulse generation are fundamental requirements for photonic quantum computation and communication. In many photonic quantum platforms, optical pulses define the temporal modes used for quantum-light generation, state preparation, routing, feed-forward operations, and measurement. These functions are central to integrated quantum photonic systems, where sources, reconfigurable circuits, detectors, and classical control must be combined on a scalable platform \cite{Flamini2019PhotonicQIP,Slussarenko2019PhotonicQIP,Wang2020IntegratedQuantumPhotonics,Giordani2023IntegratedPhotonicsQuantum,Labonte2024IntegratedQuantumComms,Caspani2017IntegratedSources}. In these systems, residual optical leakage in the nominally $\mathrm{"off"}$ state directly degrades state-preparation fidelity, interference visibility, switching contrast, and quantum-bit-error rate. Therefore, sharp optical pulses with high extinction ratio are not merely desirable, but are a basic requirement for scalable quantum photonic systems. Despite this importance, scalable generation of high-extinction, low-leakage optical pulses remains challenging in integrated photonics, as it requires a low-loss and reliable electro-optic platform that can operate at low drive voltage, remain compatible with cryogenic environments, preserve pulse fidelity without temporal distortion, and support a broad range of modulation frequencies. Thermo-optic silicon switches and phase shifters are widely used for low-loss reconfiguration in silicon photonic circuits, including large-scale switch fabrics \cite{Harris2014ThermoOptic,Suzuki2017Broadband8x8,Suzuki2018LowInsertion32x32}. However, their operation is fundamentally limited by thermal diffusion, resulting in kHz--hundreds-of-kHz modulation bandwidths and microsecond-to-millisecond switching times \cite{Harris2014ThermoOptic,Suzuki2017Broadband8x8,Suzuki2018LowInsertion32x32}. Therefore, thermo-optic switches are well suited for static trimming and slow reconfiguration, but are not suitable for high-speed optical pulse generation or fast quantum feed-forward control. Silicon photonics has been widely explored for quantum photonic circuits \cite{Wang2020IntegratedQuantumPhotonics,Flamini2019PhotonicQIP,Silverstone2016SiliconQuantumPhotonics}. However, crystalline silicon is centrosymmetric and therefore lacks a native linear Pockels effect \cite{Reed2010SiliconModulators,Alexander2018PockelsSiN}. High-speed silicon modulators therefore rely predominantly on free-carrier plasma dispersion, where carrier injection or depletion changes both the refractive index and absorption, leading to free-carrier absorption and intrinsic phase--amplitude coupling \cite{Soref1987ElectroopticalSilicon,Reed2010SiliconModulators,VanThourhout2021Modulators}. While such loss can often be tolerated in classical communication systems, where optical amplification is available, it is highly detrimental in quantum photonic systems because photon loss directly reduces success probability and cannot be compensated without adding noise or destroying the quantum state. This limitation has motivated heterogeneous silicon-photonic approaches, including Si--BTO platforms for low-power Pockels-based electro-optic switching. PsiQuantum's recent silicon-photonic quantum computing platform, for example, identifies low-loss SiN components and BTO electro-optic phase shifters as next-generation technologies for scalable, fast, low-loss switching networks. Nevertheless, such platforms remain comparatively complex and can still require auxiliary tuning or trimming elements, such as thermal phase shifters, which increase heat load in cryogenic systems \cite{PsiQuantum2025ManufacturablePlatform,Eltes2019BTOSilicon,Eltes2020CryogenicBTO}. Among low-loss electro-optic material platforms, AlN, LN, and LT are particularly attractive. AlN enables low-loss, CMOS-compatible photonic circuits \cite{Liu2023AlNPhotonicsReview,Xiong2012LowLossAlN}, but its Pockels coefficients are relatively small, with typical reported values near $\sim 1~\mathrm{pm/V}$, leading to substantially larger drive voltages or device lengths compared with LN or LT \cite{Zhu2016AlNPhaseShifter,Liu2023AlNPhotonicsReview}. BTO offers a very large effective Pockels response, but practical BTO-integrated photonic circuits often face challenges associated with heterogeneous integration, optical loss, material uniformity, ferroelectric domain control, and long-term bias stability \cite{Eltes2019BTOModulator,Tao2024TransferredBTO,Li2024PockelsModulatorsReview}. 
\begin{figure*}[t!]
    \centering
    \includegraphics[width = 0.85\textwidth]{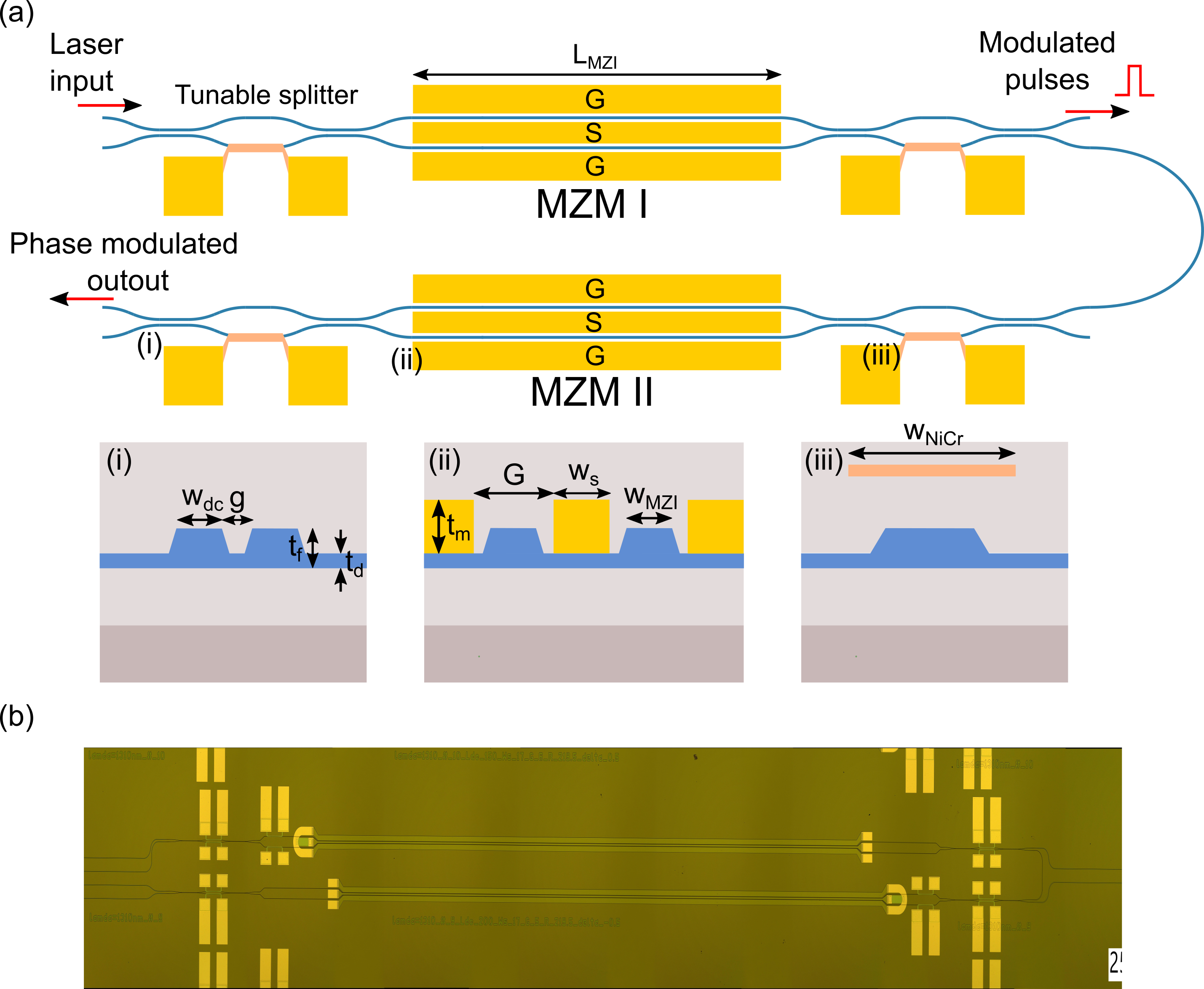}
    \caption{Schematic of the TFLT MZM modulator with tunable splitters. Cross-sectional views are shown for different device sections: (i) directional coupler, (ii) MZI section with traveling-wave coplanar ground–signal–ground (GSG) electrodes, and (iii) thermal tuner with NiCr heaters. The device parameters are, $\mathrm{w_{dc}=0.8\mu m, g=0.8\mu m, t{m}=1.0\mu m, G=5.0\mu m, w_{s}=17.5\mu m, w_{MZI}=1.6\mu m, w_{NiCr}=10\mu m}$. (b) Optical microscope image of the MZM modulators terminated with on-chip $\mathrm{50,\Omega}$ resistors.}
    \label{fig0}
\end{figure*}
In contrast, thin-film lithium niobate (TFLN) combines low optical propagation loss with a large, ultra fast Pockels effect, making it an appealing platform for high-speed electro-optic modulation \cite{zhang2017monolithic,wang2018integrated,zhu2021integrated,kharel2021breaking,Zhang2021IntegratedLNModulators}. However, for quantum pulse generation and switching, high bandwidth alone is insufficient. TFLN devices can exhibit charge-related bias drift, photorefractive and pyroelectric memory effects, and long-lived transient responses that distort the optical waveform after switching events \cite{Ren2025PhotorefractiveMemory,Shen2024UltraHighERPulses}. Such dynamics can produce pulse tails or incomplete return to the off state, limiting the achievable extinction ratio in practical pulse-mode operation. Recent experiments have shown that ultra-high-extinction optical pulses on TFLN require careful engineering of the electrical driving waveform to compensate a relaxation-tail response attributed to interface-carrier dynamics \cite{Shen2024UltraHighERPulses}. Although effective in a single device, this approach is difficult to scale to large quantum photonic circuits, because the required compensation waveform can depend on device geometry, optical power, fabrication conditions, temperature, and wafer-scale material variation. Therefore, a platform that can intrinsically generate sharp, high-extinction optical pulses without device-specific waveform pre-compensation would represent a major advantage for scalable quantum photonic computation and communication systems.

In this work, we experimentally demonstrate sharp and broadband-duration optical pulse generation in thin-film lithium tantalate (TFLT). By using tunable splitters to balance the splitting ratio of a Mach--Zehnder interferometer, we generate high-extinction optical pulses with no measurable waveform distortion and an extinction ratio approaching $40\,\mathrm{dB}$, limited primarily by scattered light. We then systematically investigate the physical origin of pulse tails in TFLN and show that these long-lived transients are absent in TFLT. Temperature-dependent current-voltage (I-V) measurements reveal a larger activation energy for defect-mediated charge transport in TFLT than in TFLN, with extracted values of $\mathrm{E_{a,TFLT}=1.306\,eV}$ and $\mathrm{E_{a,TFLT}=0.894\,eV}$, respectively. The corresponding pre-exponential factors are remarkably similar for the two platforms, with $A = 138.95~\mathrm{S}$ for TFLT and $A = 123.21~\mathrm{S}$ for TFLN, indicating that the reduced charge transport in TFLT is primarily governed by its larger activation barrier rather than by differences in the prefactor. Because thermally activated transport scales as $A\exp(-E_a/k_B T)$, the higher activation energy in TFLT corresponds to a strong suppression of mobile charge dynamics at room temperature, preventing the slow relaxation processes that lead to pulse tails and waveform distortion in TFLN. Consistent with this picture, we generate distortion-free optical pulses with widths up to $1\,\mathrm{s}$, confirming that high-extinction pulse generation in TFLT is not limited by long-lived defect-charge effects. Our results position TFLT as a robust monolithic electro-optic platform for scalable integrated quantum photonic systems. 

\section{Device architecture}

Fig.\ref{fig0} shows the schematic of the TFLT device used in this experiment, along with an optical microscope image. The device consists of two cascaded Mach-Zehnder modulators (MZM) connected in series. Each modulator consists of three MZIs, the first and the third MZIs act as tunable splitters where the beam splitting can be be tuned by using therm-optic (TO) heaters. The second MZI of each modulator has traveling wave co-planar waveguide (CPW) electrodes for optical modulation. For both modulators, the signal width of the electrode, $\mathrm{w_{s}=17.5\,um}$, and the gap between the ground and the signal electrode, $\mathrm{G=5.5\,um}$. The details of the fabrication process can be found in our previous work \cite{sayem2026high_1um}. 

A conventional direct-detection method utilizing a photodetector (PD) and a digital oscilloscope provides a dynamic range of approximately $25$ dB due to vertical quantization limits, low-frequency detector noise, and scattered ambient light. This is typically sufficient for resolving the strong relaxation dynamics of high extinction ratio (ER) optical pulses in TFLN platforms. However, to be able to capture the ultra-high ER dynamics of our TFLT modulators, we implement a cascaded modulation scheme paired with an electrical spectrum analyzer (ESA) operating in zero-span mode. This approach bypasses low-frequency noise to achieve a dynamic range that can, in principle, exceed $80$ dB. In this technique, a continuous-wave (CW) laser is routed into the first TFLT modulator to imprint the temporal pulse profile $A(t)$, yielding an optical field:$$E_1(t) = A(t)\exp(i\omega t)$$where $\omega$ is the optical angular frequency. This pulsed field is subsequently passed through a second modulator driven by a radio-frequency (RF) tone at frequency $\Omega$.The resulting output field after the second modulator $E_2(t)$ can be expressed as:
\begin{align*} E_2(t) &= A(t)\exp(i\omega t)\cos\left(\frac{\pi}{4} + \frac{\beta}{2}\cos(\Omega t)\right) \\ &\approx \frac{1}{\sqrt{2}}A(t)\exp(i\omega t)\left[1 - \frac{\beta}{2}\cos(\Omega t)\right] \end{align*}
where $\beta$ is the modulation index. The photodetector captures the intensity of this field, which scales as:$$\vert{}E_2(t)\vert{}^2 \approx \frac{1}{2}\vert{}A(t)\vert{}^2 [1 - \beta\cos(\Omega t)]$$This creates an RF beat note at $\Omega$ that is directly proportional to the time-dependent pulse power profile $\vert{}A(t)\vert{}^2$. By configuring the ESA to zero-span mode centered at $\Omega$, the deep temporal profile of the pulse can be mapped with high dynamic range, effectively mapping out the complete $\vert{}A(t)\vert{}^2$ profile without low-frequency noise degradation.

\begin{figure*}[ht]
    \centering
    \includegraphics[width = .7\textwidth]{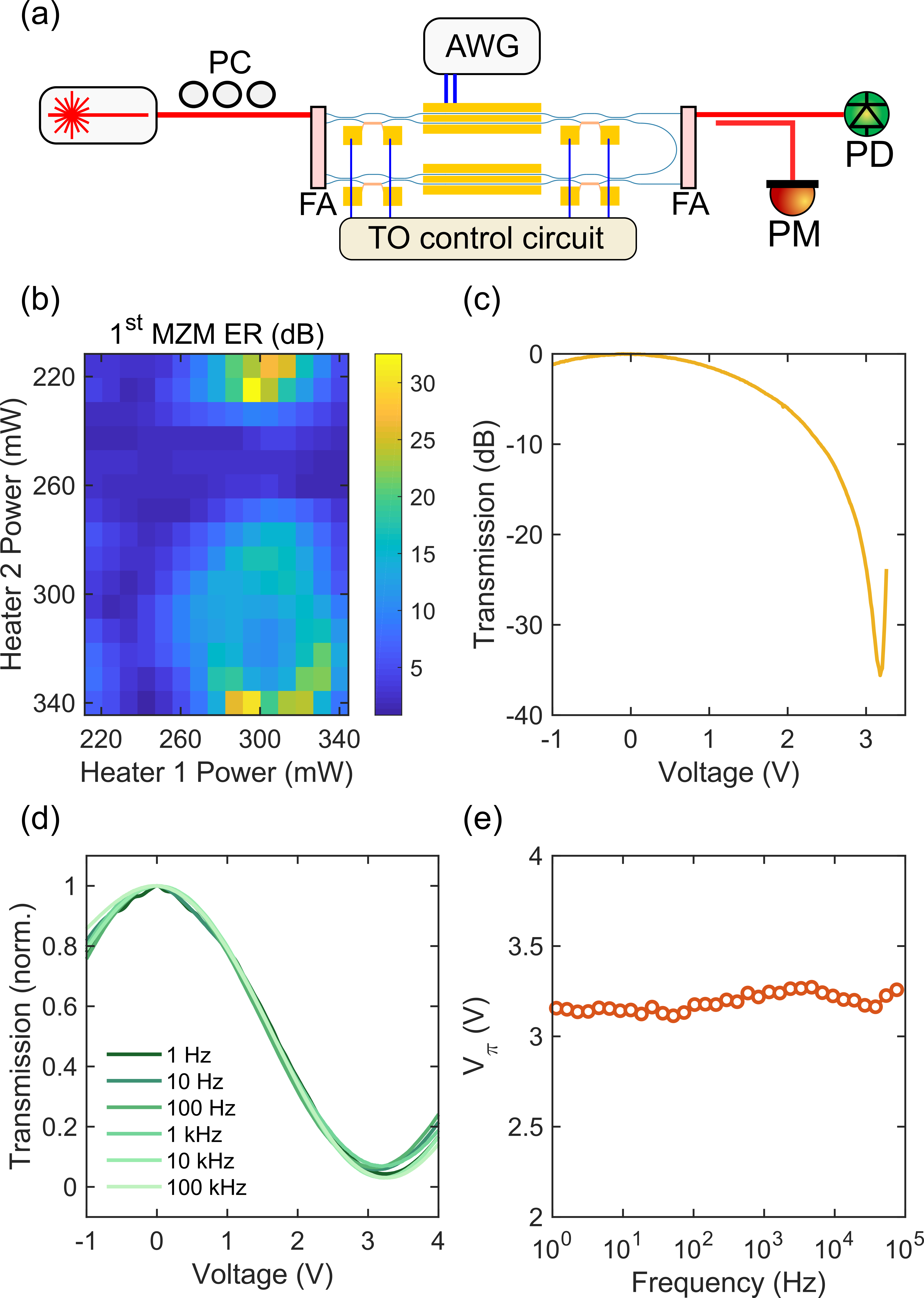}
    \caption{ (a) Experimental setup used to characterize the individual MZZMs, (b) ER of the first MZI modulator as a function of the heater's powers, (c) Transmission as a function of applied voltage with optimized heater bias. (d) Transmission as a function of applied voltage with different sweep frequencies of the voltage drive. (e) Extracted $\mathrm{V_{\pi}}$ as a function of the RF drive frequency.} 
    \label{fig1}
\end{figure*}

\section{Experimental results}
The experimental setup used to optimize the top MZM is shown in Fig.~\ref{fig1}(a). Light from a tunable laser (Santec TSL-570) is coupled into the device under test (DUT) using a fiber-based polarization controller (PC) followed by a fiber array (FA) composed of ultra-high numerical aperture (UHNA-7) fibers to better match the fundamental TE mode of the TFLT waveguides. The output light is collected using a second UHNA fiber array and directed through a fiber splitter to a slow optical power meter (PM) and a high-speed photo-detector (PD, Thorlabs PDA10CS2). The photo-detector output is recorded using either an oscilloscope or a data acquisition unit (DAQ). Before pulse measurements, we optimize the tunable waveguide beam splitters to maximize the extinction ratio of the first MZM, MZM-I. In this configuration, the upper-left input port and upper-right output port are used as the modulator input and output, respectively. Figure~\ref{fig1}(b) presents a two-dimensional map of the extinction ratio as a function of the heater powers applied to the tunable splitters. Figure~\ref{fig1}(c) shows the optical transmission as a function of applied voltage where we use the slow power meter (PM) with better dynamic range to measure the transmission. Figure~\ref{fig1}(d) compares the transmission--voltage characteristics for RF drive frequencies from $1\,\mathrm{Hz}$ to $100\,\mathrm{kHz}$, while Fig.~\ref{fig1}(e) shows the corresponding extracted $V_{\pi}$ values. The nearly unchanged modulation curves and $V_{\pi}$ values across this frequency range demonstrate stable, frequency-independent electro-optic operation.

\begin{figure*}[ht!]
    \centering
    \includegraphics[width = 0.70\textwidth]{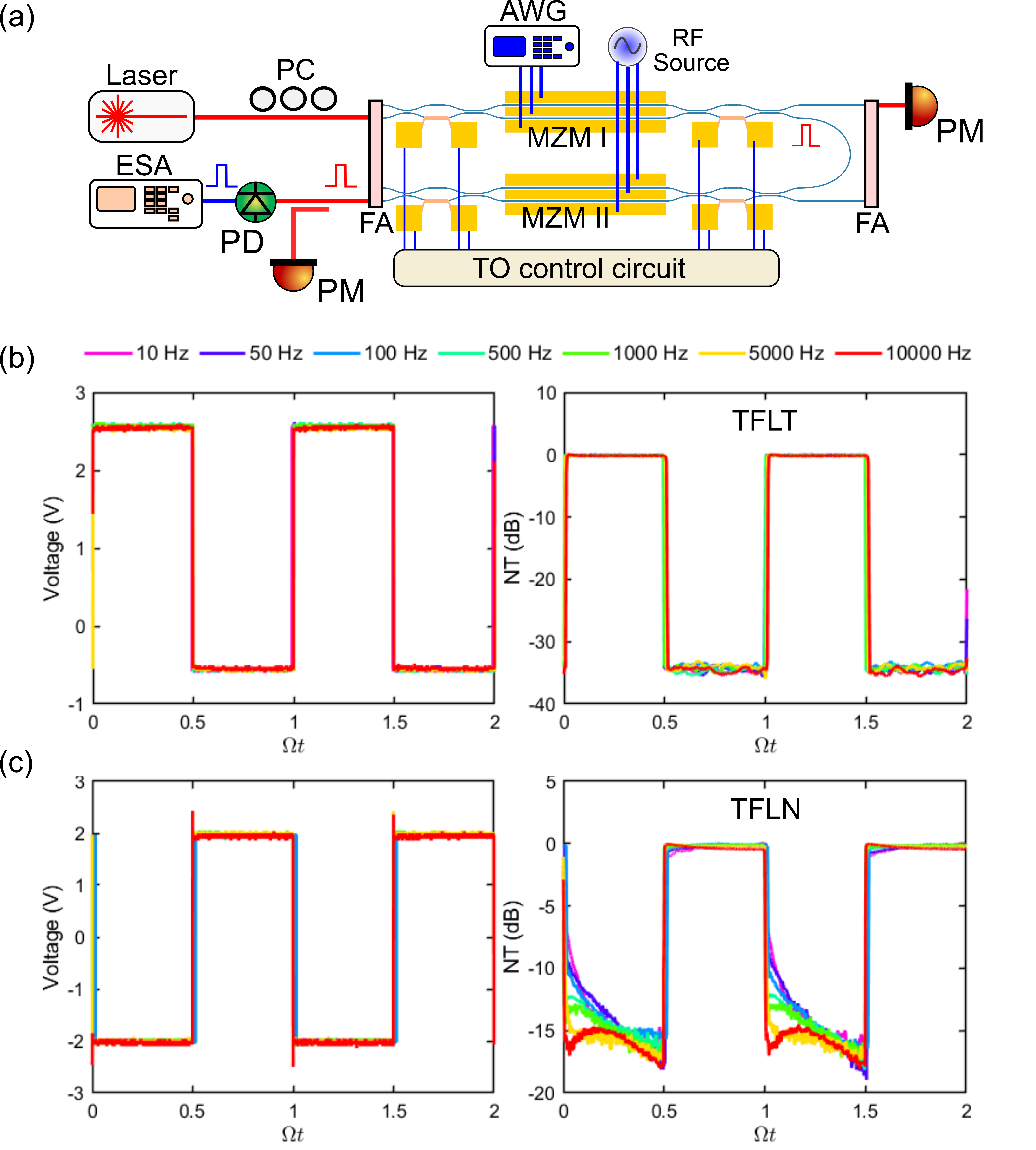}
    \caption{(a) Schematic of the measurement setup for optical pulse generation in TFLT. Applied voltage waveforms and normalized transmission (NT) of the output optical pulses from the MZM as a function of time at different drive frequencies for (b) TFLT and (c) TFLN.}
    \label{fig2}
\end{figure*}

\begin{figure*}[ht!]
    \centering
    \includegraphics[width = 0.70\textwidth]{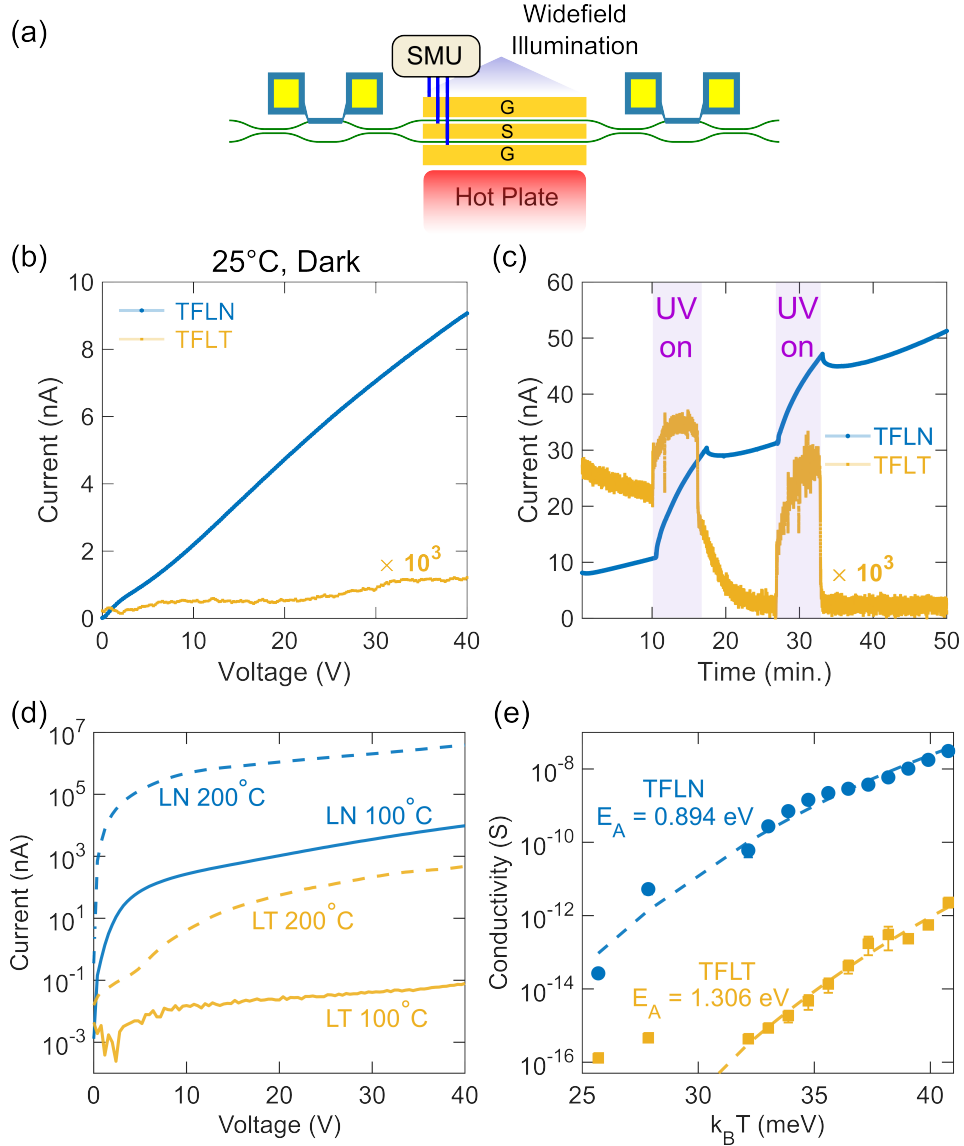}
    \caption{(a) Schematic of the experimental setup illustrating the widefield UV illumination of the Mach-Zehnder modulator (MZM) and the electrical probing configuration using a Source Measure Unit (SMU) connected to the ground-signal-ground (GSG) electrodes. The device is placed on a temperature-controllable hot plate. (b) Current-voltage ($I-V$) characteristics for the upper-SiO$_2$ clad thin-film lithium tantalate (TFLT; $4\mu\text{m}$ gap, $7\text{mm}$ length) and thin-film lithium niobate (TFLN; $5\mu\text{m}$ gap, $6\text{mm}$ length) devices at room-temperature ($25^\circ\text{C}$) without any UV illumination.  (c) Transient current response under a constant $40\text{V}$ bias. The shaded regions indicate periods of widefield UV illumination, Note that the TFLT current is scaled by a factor of $10^3$ for visual comparison in both (b) and (c). (d) $I-V$ characteristics of TFLT and TFLN devices in the dark at elevated temperatures. (e) Sheet conductivity ($\sigma$) as a function of thermal energy ($k_B T$), where $k_B$ is the Boltzmann constant and $T$ is the absolute temperature. The temperature dependence follows the Arrhenius relation $\sigma = A e^{-E_a / k_B T}$ shown in dashed line.}
    \label{fig3}
\end{figure*}
After optimizing the first modulator, MZM-I, we used the cascaded modulator, MZM-II, to resolve the temporal dynamics of the high-extinction pulses. Figure~\ref{fig2}(a) shows a schematic of the measurement setup used to characterize the pulse dynamics, which is similar to that shown in Fig.~\ref{fig1}(a). In this configuration, light is coupled into the lower input port of MZM-I (top), and the output light is collected from the lower output port of MZM-II (bottom) using another channel of the fiber array. A thermo-optic bias-control circuit is used to set MZM-I at its maximum-extinction operating point and MZM-II at its quadrature point. The extinction ratio of MZM-II as a function of the heater powers applied to the two heaters of MZM-II are very similar to MZM-I. In this case, light is coupled to MZI-II through MZM-I when MZM-I is not being modulated. We drive MZM-II with a RF tone at $\mathrm{\Omega_{RF}=5\,GHz}$. The output of MZM-II is sent to a $15\text{\,GHz}$ photoreceiver and then to an electronic spectrum analyzer (ESA) which is set to zero span mode at $5\text{\,GHz}$ and time-sweep configuration. Fig.\ref{fig2}(b) shows the voltage pulses and the corresponding output pulses from the TFLT modulator. More than 35\,dB extinction pulses can be generated with wide range of frequencies as low as 10\,Hz without any distortion. In our experiment, the measured extinction ratio is primarily limited by scattered light coupling from the input channel of the fiber array to the output channel and the dynamic range of the PD and the ESA. Using a strong local oscillator can further resolve higher extinction \cite{shen2023ultra}. The measured pulse rise and fall times are limited by the $30$\,MHz arbitrary waveform generator (AWG) used in the experiment, whereas the modulator itself has an electro-optic bandwidth of $60$\,GHz. Currently, the ER is limited by the scattered light, detector floor and, laser power limit, With improved fiber-to-chip coupling schemes, such as efficient grating couplers or edge couplers based on double inverse tapers, we expect the extinction ratio to exceed 35\,dB.

We next compare these results with those obtained from a TFLN modulator. For the TFLN device, we use a simple modulator design in which a 1$\times$2 multi-mode interferometer (MMI) is used as the beam splitters instead of a tunable directional coupler. Consequently, the DC extinction ratio of the TFLN modulator is limited to 18\,dB. Nevertheless, even at this relatively low extinction ratio, pronounced long tails are observed in the output pulses from the TFLN modulator, as shown in Fig.~\ref{fig2}(c). As the strong relaxation tails of TFLN modulators are readily observable even at low extinction levels, a second modulator to enhance the dynamic range is unnecessary. This behavior is consistent with previous reports \cite{Shen2024UltraHighERPulses} and poses a major challenge for generating sharp, high-extinction optical pulses over a broad range of repetition frequencies, particularly for quantum photonic applications. 

We investigate the stark differences between the electrical response of TFLT and TFLN modulators using simple current--voltage (I--V) measurements. For this measurement, we use identical, non-terminated modulators fabricated on both TFLT and TFLN platforms. The CPW electrode is biased using a source-measure unit (SMU: Keithley 2405), as shown in Fig.~\ref{fig3}(a). In Fig.~\ref{fig3}(b), we plot the measured current as a function of applied voltage at room temperature under dark conditions, with only regular room light present. The TFLN device exhibits nearly four orders of magnitude higher current than the TFLT device. Roughly four orders of magnitude lower sheet conductivity is observed in TFLT ($8.7 \times 10^{-18}$ S/$\square$) compared to TFLN ($9.4 \times 10^{-14}$ S/$\square$). We then measure the current as a function of time under a constant bias of 40\,V while periodically illuminating the modulators with a UV diode source ($\lambda = 365\,\mathrm{nm}$, source fluence of $240\,\mathrm{mW/cm^2}$, and source-to-device distance of $\sim 6\,\mathrm{mm}$). Flood illumination is applied to both TFLN and TFLT devices. Although this measurement is not intended to quantify the absolute photoconductivity of the platforms as the whole length of the device is not uniformly illuminated in a controlled way, it provides a qualitative comparison of how UV exposure modulates their respective leakage currents. A significant increase in current is observed for the TFLN modulator when the UV light is turned on. In contrast, the TFLT modulator shows only a slight increase in current, which remains orders of magnitude lower than that of the TFLN device. We further perform temperature-dependent I--V measurements. Fig.~\ref{fig3}(c) shows the I--V characteristics of the TFLN and TFLT modulators at elevated temperatures of $100^{\circ}\mathrm{C}$ and $200^{\circ}\mathrm{C}$. While both materials show increased current with temperature, TFLT maintains significantly lower leakage across the entire measured range. From these I--V curves, we extract the conductivity of both platforms at different temperatures. We plot the conductivity, $\sigma$, as a function of thermal energy, $k_{\mathrm{B}}T$ in Fig.~\ref{fig3}(d), and fit the temperature dependence using an Arrhenius-type model $\sigma = A e^{-E_a / k_B T}$. The pre-exponential factor $A$ represents the intrinsic limit of conductivity as the temperature approaches infinity ($T \to \infty$), where the exponential "barrier" term effectively drops to unity and is generally proportional to the product of the concentration of available charge carriers and their mobility.  The extracted pre-exponential factors ($A$) are remarkably similar for both platforms-$138.95\text{ S}$ and $123.21\text{ S}$ for TFLT and TFLN, respectively. It indicates that the underlying mechanism for transport is likely similar in both materials. Although the pre-exponential factors are comparable for the two platforms, the activation energies differ significantly. The significantly higher activation energy of TFLT ($E_a = 1.306\text{ eV}$) compared to TFLN ($E_a = 0.894\text{ eV}$) indicates a much larger energy barrier for charge carrier activation, which fundamentally drives the orders-of-magnitude reduction in dark conductivity observed in the tantalate platform. These results reveal that the reduced leakage current in TFLT is not merely a consequence of geometric or contact-related differences, but is instead rooted in the intrinsically larger activation barrier for charge transport in the tantalate platform. These findings not only explain the distortion-free pulses observed in this work, but also provide direct experimental support for the DC-bias stability previously reported in TFLT modulators.



\bibliography{Reference}


\end{document}